\documentstyle[11pt,newpasp,twoside,epsf]{article}
\def\edcomment#1{\iffalse\marginpar{\raggedright\sl#1\/}\else\relax\fi}
\reversemarginpar

\begin{document}
\title{Accretion disk assembly and survival during the disruption of a neutron star by a black hole}
\author{Enrico Ramirez--Ruiz}
\affil{Institute of Astronomy, Madingley Road, Cambridge, CB3OHA, U.K}
\author{William H. Lee}
\affil{Instituto de Astronom\'{\i}a, 
Universidad Nacional Aut\'{o}noma de M\'{e}xico, Apdo. Postal 70-264,
Cd. Universitaria, M\'{e}xico D.F. 04510}

\begin{abstract}
We study the formation of accretion disks resulting from dynamical
three dimensional binary coalescence calculations, where a neutron
star is tidally disrupted before being swallowed by its black hole
companion. By subsequently assuming azimuthal symmetry we are able to
follow the time dependence of the disk structure for a few
tenths of a second. Although the disruption of a neutron star leads to
a situation where violent instabilities redistribute mass and angular
momentum within a few dynamical timescales, enough gas mass remains
in the orbiting debris to catalyse the extraction of energy from the
hole at a rate adequate to power a short-lived gamma ray burst.
\end{abstract}

\section{Short-lived mysteries}
Almost 5 years after astronomers first discovered their telltale
afterglows, the brief, ultrabright flashes of high-energy radiation
known as cosmic gamma-ray bursts (GRBs) are still among the most
mysterious phenomena in the universe. Their energy output has to be of
the order $10^{51}-10^{54}$ erg~s$^{-1}$, larger than that of any
other type of source. Most researchers agree that the most common GRBs
-- those that last between about a second and a minute -- signal the
catastrophic collapse of massive, rapidly rotating stars into black
holes (see M\'esz\'aros 2001 for a review).  The precise details of
their origin, however, are unknown; and the nature of bursts that
flash in less than a second is anybody's guess. These bursts, which
account for about one-third of all observed GRBs, differ markedly from
the long ones not only in duration but also in having a larger
proportion of high-energy gamma rays in their energy distribution.
The most widely favoured and conventional possibility is that they
result either from the merger of two neutron stars or of a neutron
star and a black hole (e.g. Narayan et al. 1992).

\begin{figure}
\plotfiddle{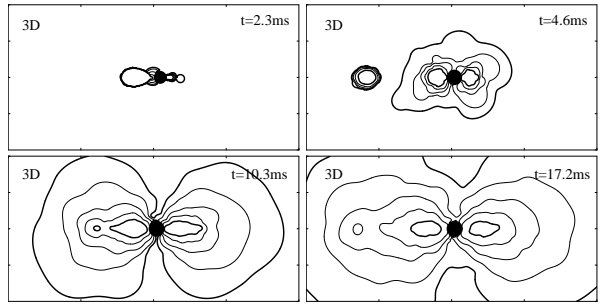}{7cm}{0}{39}{39}{-100}{68}
\vspace{-3.5cm}
\caption{Sequence of images of the tidal disruption of a neutron star
by a black hole in a close binary, showing density contours in a
meridional slice (3D calculation). All contours are logarithmic and
equally spaced $\log \rho = 8 - 12$ in cgs units. Bold contours are
plotted at $\log \rho = 8, 12$. Each box size is 400 km $\times$ 200
km.}
\end{figure}

When a neutron star is tidally disrupted by a black hole, the total
angular momentum is large enough for the star not to be swallowed
immediately. The expected outcome, after a few milliseconds, would
therefore be a spinning black hole, orbited by a torus of
neutron-density matter. An acceptable model thus requires that the
surrounding torus should not completely drain into the hole, or be
otherwise dispersed, on too short a time scale (Rees 1999). The key
issue is then how long a sufficient amount of this matter survives to
power a burst, and it is to this problem that we have turned our
attention.

\section{Coalescing odd couples}
It has been our objective to investigate the outcome of the
coalescence of a black hole with a neutron star, to find out to what
extent the neutron star is tidally disrupted and, in particular, to
determine if an accretion structure does form around the black hole as
a result of the encounter. Using a three-dimensional smooth particle
hydrodynamics (SPH) code we have simulated the last stages of binary
evolution of a black hole and a neutron star, when the components are
separated by a few stellar radii. The gravitational radiation
waveforms as well as the gravitational radiation luminosity are
calculated in the quadrupole approximation; and the neutron star is
modeled with a stiff polytropic equation of state (the reader is
referred to Lee 2001 for further details). Given that tidal locking is
not expected in these systems (e.g. Bildsten \& Cutler 1992), we have
used initial conditions that correspond to irrotational binaries in
equilibrium, approximating the neutron star as a compressible triaxial
ellipsoid.

\begin{figure}
\plotfiddle{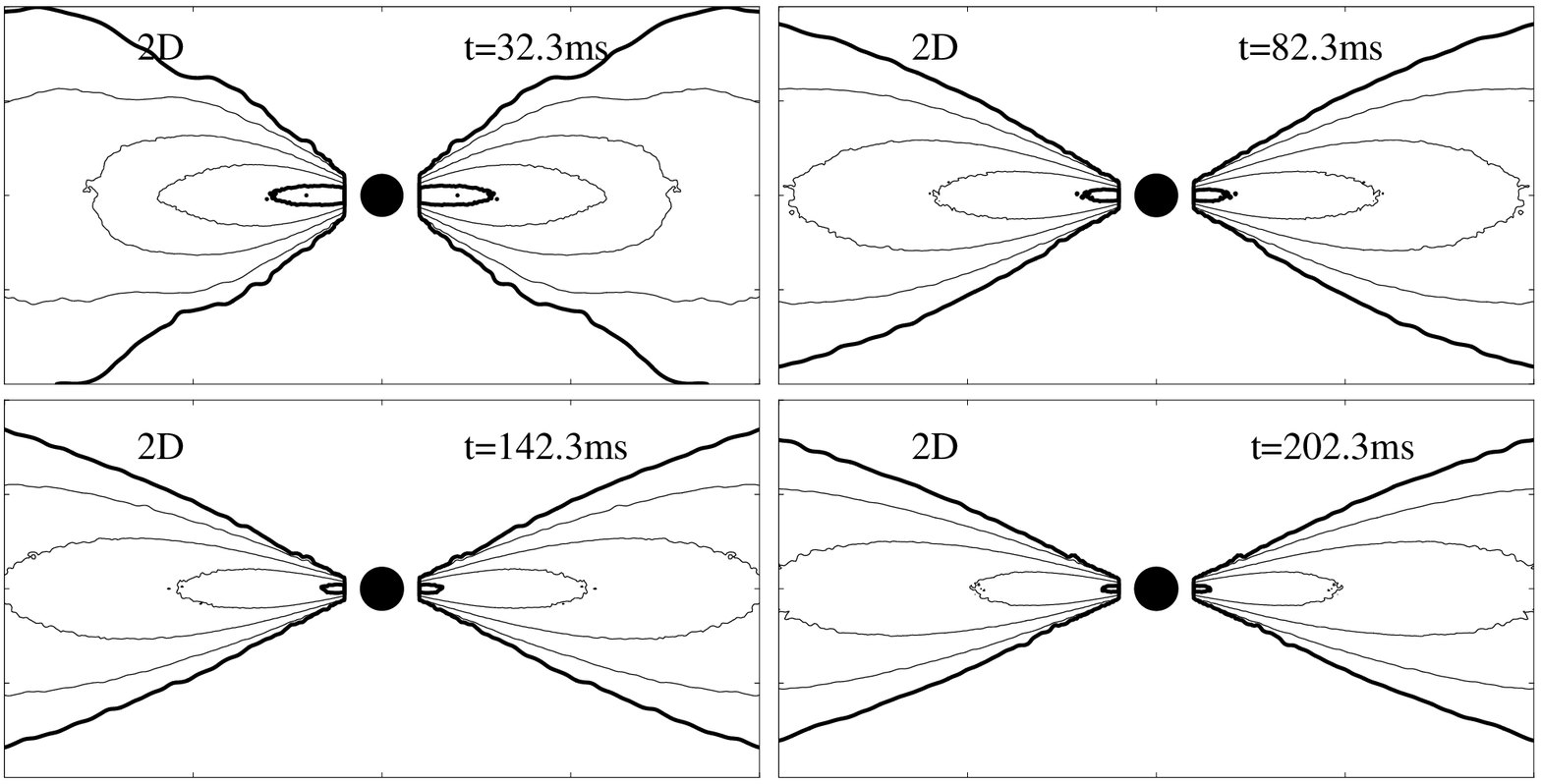}{7cm}{0}{40}{40}{-40}{64}
\plotfiddle{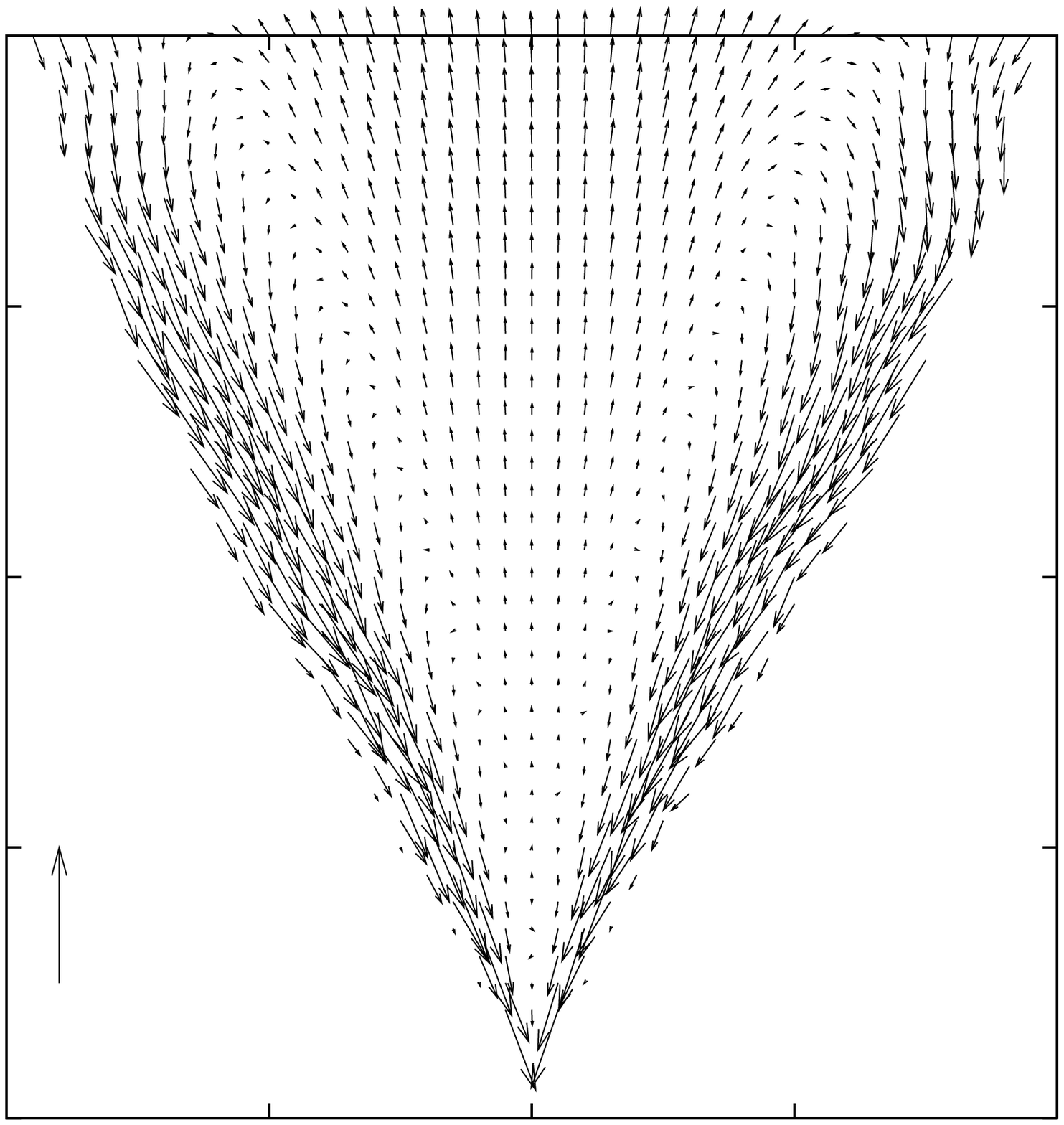}{7cm}{-90}{30}{25}{-210}{436}
\vspace{-10.5cm}
\caption{Dynamical evolution of a massive accretion disk around a
stellar mass black hole in two dimensions (azimuthal symmetry). {\it
Left:} Velocity field at $t=82.3$ ms. The vector at top left has
$v=5\times 10^{8}$cm~s$^{-1}$. {\it Right:} logarithmic density
contours as those shown in Fig.~1.}
\end{figure}
The 3D dynamical simulations are begun when the system is on the verge
of initiating mass transfer, and followed for approximately 22 ms. The
decrease in binary separation leads to Roche lobe overflow on an
orbital time-scale. A stream of gas forms at the inner Lagrange point,
transferring matter from the neutron star to the black hole. At the
same time, the star is tidally stretched and extends away from the
black hole through the external Lagrange point. Figure 1 shows the
density contours in a meridional slice at various times during the
simulation.

As the accretion stream winds around the black hole, it collides with
itself and forms a torus, while the gas thrown out through the outer
Lagrange point forms a long tidal tail. The orbiting debris that is
formed around the black hole is not initially azimuthally symmetric,
but shows a double ring structure. This appears as the gas that passes
through periastron near the black hole overshoots the circular orbit
that would correspond to the specific angular momentum it contains,
forming an outer ring. It then falls back towards the black hole and
encounters the rear of the accretion stream. The subsequent collision
tends to circularize the orbit of the fluid, and also pushes it to the
inner ring, closer to the black hole. The structure of the outer ring
and the bulk of the disrupted star continue their orbital motion on
opposite sides of the black hole. At late times, the disc becomes more
azimuthally symmetric.

\section{Survival of the orbiting debris}

A question which has remained largely unanswered so far is what
determines the characteristic duration of bursts, which can extend to
tens, or even hundreds, of seconds. This is of course very long in
comparison with the dynamical or orbital time scale. The disruption of
a neutron star is almost certain to lead to a situation where violent
instabilities redistribute mass and angular momentum within a few
dynamical time scales. A key issue is then the nature of the surviving
debris after these violent processes are over.

So far, though, attempts to calculate such an evolution have run up
against the lack of multi--dimensional, time--dependent
simulations. There is now a stronger motivation to develop models in
fuller detail. This paper outlines the behavior of the accretion
structures on timescales that are much longer than the dynamical
one. By assuming azimuthal symmetry we are able to map the output from
the 3D calculation described above to 2D, and thus follow the time
dependence of the disk structure on timescales that are comparable to
the durations of short bursts (in this case for an additional
200~ms). We use Newtonian physics, an ideal gas equation of state, and
solve the equations of viscous hydrodynamics assuming an $\alpha$
law. All the energy dissipated by the physical viscosity is radiated
away in neutrinos (Lee \& Ramirez-Ruiz 2002).

We find that meridional circulations are promptly established, whose
structure depends mainly on the value of $\alpha$. There is an
important motion of fluid from the inner regions of the disks to large
radii, along the equatorial plane. The flow is directed toward the
accreting black hole along the surface of the disk and in the
equatorial region at small radii (Fig. 2 shows the evolution for
$\alpha=0.1$). The disks remain thick ($H/R \sim 0.5$) throughout the
dynamical evolution, due to their large internal energy, with
accretion rates on the order of 1 $M_\odot {\rm s}^{-1}$. The maximum
densities decrease during our calculations, as there is no external
agent feeding the disks, but remain at $\sim 10^{12} {\rm
g\;cm}^{-3}$, with corresponding internal energy densities $\sim {\rm
few}\; \times 10^{30} {\rm ergs\; cm}^{-3}$.

There are several instabilities that can affect massive accretion
disks dynamically and shorten their lifetimes considerably, compared
with the viscous timescale. These instabilities can be virulent in a
torus where the specific angular momentum is uniform throughout, but
are inhibited by a spread in angular momentum. In a torus that was
massive and/or thin enough to be self-gravitating, bar-mode
gravitational instabilities could lead to further redistribution of
angular momentum and/or to energy loss by gravitational radiation
within only a few orbits. Whether a torus of given mass is dynamically
unstable depends on its thickness and stratification, which in turn
depends on internal viscous dissipation and neutrino cooling (Narayan
et al. 2001).

From our simulations we draw the following conclusions. First, we find
that the central object survives the initial, violent event that
created it -- the total angular momentum is large enough for the star
not to be swallowed immediately. Second, the orbiting debris is stable
to both radial and axisymmetric perturbations. Third, even if the
evolution time scale for the bulk of the debris torus were no more
than a tenth of a second, enough may remain to catalyse the
extraction of energy from the hole at rate adequate to power a
short-lived burst. However, these simulations clearly cannot tackle
directly other relevant issues, mainly related to the evolution of the
magnetic field, and its influence on the dynamics.

Magnetic instabilities could make the disk lifetime much shorter by
effectively increasing the viscosity. The amplification of the
magnetic field may be self--limiting due to magnetic stress, which
would cause disk flaring. The properties of the expected variability
depend strongly on the details of the configuration of the disk corona
(e.g. Thompson 1994) generated by the magnetic field, which
is removed from the disk by flux buoyancy.

\end{document}